\documentclass[12pt,preprint]{aastex} 

\newcommand{\be}  {\begin{equation}}
\newcommand{\ee}  {\end{equation}  }
\newcommand{\bea} {\begin{eqnarray}}
\newcommand{\eea} {\end{eqnarray}  }
\newcommand{\rinf}{ \ensuremath{R_{\infty} }}
\newcommand{\tinf}{ \ensuremath{T_{\infty} }}
\newcommand{\rg}  { \ensuremath{R_{\rm g}  }}

\newcommand{\msun}{ \ensuremath{M_{\odot}  }}
\newcommand{\rmag}{ \ensuremath{r_{\rm m}  }}
\newcommand{\rco} { \ensuremath{r_{\rm co} }}

\newcommand{\mdot}{ \ensuremath{{\dot M}   }}
\newcommand{\frad}{ \ensuremath{f_{\rm rad}}}
\newcommand{\bbag}{ \ensuremath{{\bar B}   }}
\newcommand{\mum} { \ensuremath{\mu_{\rm m}}}
\newcommand{\pdot}{ \ensuremath{\dot P     }}

\newcommand{\obl} { \ensuremath{\Omega_{\rm BL}}}

\shorttitle{The UV-optical Excess of RX J1856}

\shortauthors{Yue, Cui \& Xu}

\begin{document}

\title{To Understand the UV-optical Excess of RX J1856.5-3754}

\author{Y. L. Yue, X. H. Cui and R. X. Xu}
\affil{School of Physics, Peking University, Beijing 100871,
China; {\tt r.x.xu@pku.edu.cn}}

\begin{abstract}

The enigma source, RX J1856.5-3754, is one of the so-called dim
thermal neutron stars. Two puzzles of RXJ1856.5-3754 exist: (1)
the observational X-ray spectrum is completely featureless; (2)
the UV-optical intensity is about seven times larger than that
given by the continuation of the blackbody model yielded by the
X-ray data. Both the puzzles would not exist anymore if RX
J1856.5-3754 is a low mass bare strange quark star, which is in a
propeller phase with a low accretion rate. A boundary layer of RX
J1856.5-3754 is suggested and modelled, from which the UV-optical
emission is radiated. Free-free absorption dominates the opacity
of the boundary layer, which results in the opacity to be high in
UV-optical but low in X-ray bands.
The star's magnetic field, spin period, as well as the accretion
rate are constrained by observations.

\end{abstract}


\section{Introduction}

Quarks (and leptons) are fundamental fermions in the standard
model of particle physics, the underlying theory of the
interaction between which is believed to be quantum chromodynamics
(QCD). Quark matter (or quark-gluon plasma) is expected in QCD,
but is not uncovered with certainty.
While physicists are trying hard to find quark matter by
ground-based colliders, astronomers are researching on the
existence and/or consequence of quark matter in the sky. Quark
(matter) stars may form by supernova explosions, which are
suggested to manifest as pulsar-like compact stars.
RX J1856.5-3754 (hereafter RX J1856) is one of the {\em
ROSAT}-discovered so-called dim thermal neutron stars. It was
thought to be actually a quark star \citep{Drake02,xu02}. But this
idea was soon questioned by \cite{Walter02} if a two-blackbody
model is assumed in order to fit the observations by {\em
Chandra}, {\em EUVE}, and {\em HST}.

Among the isolated compact objects, RX J1856 is the mostly studied
one because it is the brightest.
The parallax distance of this source is only $117\pm 12$ pc
\citep{Walter02}. Its optical counterpart, with V$\sim$26 mag, was
found \citep{Walter97}, but no radio counterpart has been
observed.
A 505-ks {\em Chandra} observation revealed a featureless spectrum
which can be well fitted by a blackbody, with apparent radius
\rinf = 4.4 km and temperature \tinf = 63 eV \citep{Burwitz03},
and no X-ray modulation being found. \citet{Ransom02} setted an
upper limit on the pulsed fraction of $\sim$4.5\% (99\%
confidence) for frequency $\lesssim$ 50 Hz and frequency
derivative $-5\times10^{10}{\rm ~Hz~s^{-1}} \leq f \leq 0 {\rm
~Hz~s^{-1}}$, whereas \citet{Burwitz03} obtained an upper limit of
1.3\% (2$\sigma$ confidence) on the pulsed fraction in the
frequency range $(10^{-3}-50)$ Hz using {\em XMM-Newton} data.

Various efforts have been made to explain the observations in the
regime of normal neutron star, such as a two-component blackbody
\citep{trumper03}, a neutron star with reflective surface
\citep{Burwitz03}, a magnetar with high kick-velocity
\citep{Mori03}, a naked neutron star \citep{tzd04}, and a surface
with strong magnetic field \citep{vanA05,pmp05}. But these models
are far away from fitting the real data with reasonable
emissivity.
Summarily, several difficulties are in the normal neutron star
models.
(1) Featureless spectrum could be hard to reproduce by a neutron
star with an atmosphere of normal matter. Though strong magnetic
field may smear out some spectral features \citep{Lai01}, the
consequent large spin down luminosity is not observed.
(2) Special geometry is needed to explain the no-modulation
observation --- the pulsar should be aligned or we are situated at
the star's polar direction.
(3) A normal neutron star of radius 17 km in the two-component
model requires a very low-mass about $0.4\msun$. The mechanism to
form such a normal neutron star is still unknown.

Alternatively, the X-ray observation alone may be understood by
assuming RX J1856 to be a low-mass bare strange quark star, and
\citet{zxz04} fitted well the X-ray data with a phenomenological
spectral emissivity in a solid quark star model.
However, the UV-optical observations revealed a seven times
brighter source comparing to that derived from X-ray observation.
What's missed here?
To overcome these difficulties, we present a model under low-mass
quark star regime. A boundary layer around RX J1856 is proposed in
this paper, which is optically thick for UV-optical radiation but
is optically thin for X-ray radiation. We can obtain consistency
with the observational data, as well as strong constrains on the
star's magnetic field strength and spin period, through this
approach.

\section{The model}

We consider the star to be a bare quark star which would be
indicated by the featureless spectrum from the 505-ks {\it
Chandra} data \citep{xu02}. Since a bare quark star have quark
surface instead of atmosphere, it could reproduce a Planck-like
spectrum, rather than a spectrum with atomic lines \citep{xu03}.
Additionally, the small apparent radius ($\rinf$ = 4.4 km)
observed in X-ray band could be another hint. For these reasons, a
bare quark star hypothesis might be reasonable. The following
discussions are under this hypothesis.

The object, RX J1856, in our model consists of two components: (1)
a central bare quark star which radiates X-ray photons; (2) a
boundary layer at the magnetic radius $\rmag$ (Alfv\'en radius)
with a quasi thermal spectrum.
There are also two free parameters in our model: (1) \mum, the
magnetic moment per unit mass of the quark star; (2) \mdot, the
accretion rate. Strong constrains on these two parameters will be
given by observations in this model.
Accreted matter should be stopped at the magnetosphere and form a
boundary layer. According to following calculations, the boundary
layer is optically thin for the central X-ray emission but thick
for UV-optical emission because the cross section is much smaller
for X-ray photons than UV-optical ones. By this model, we can fit
the observations in UV-optical and X-ray bands. The details are
below.

\subsection{The star}

For a low-mass quark star, the internal density is almost
homogenous \citep{Alcock86}. The star's mass and density can be
well approximated as
\be M \simeq {\frac{4}{3}}\pi R^3 \rho ~~{\rm and}~~  \rho \simeq
4 \bbag, \label{M-R} \ee
respectively, where the bag constant is suggested to be $ \bbag =
(60\sim 100) ~{\rm MeV~ fm^{-3}}
      = 1.07 \sim 1.96\times10^{14}~ {\rm g~ cm^{-3}}.
$
A median bag constant, $\bbag = 1.5\times10^{14}~ {\rm g~
cm^{-3}}$, \rinf = 4.4 km and \tinf = 63 eV are adopted in this
paper. Considering general relativistic effects ($\rinf =
R/(1-\rg/R)^{1/2}$), we have $R$ = 4.3 km, $M = 0.097\msun$, and
$\rg =2GM/c^2=$ 0.29 km. Since $M$ is much less than $1\msun$, the
general relativistic effects are rather small.
The magnetic moment $\mu$ of the star could be expressed as
\citep{xu05}
\be \mu = \frac{1}{2}B R^3 = M \mum. \label{B-mum}\ee
This is a relatively free parameter because it spans a large
range, $(10^{-4}\sim 10^{-6})~ {\rm G ~cm^{3} ~g^{-1}}$
\citep{xu05}. We use $\mum$ as a free parameter which will be
constrained strongly.

\subsection{The boundary layer}

Since we have never observed any accretion powers, such as strong
X-ray emission and X-ray burst, the accretion rate of the star
should be very low. Hence we consider ADAF (advection dominated
accretion flow) model for the accretion
\citep[e.g.,][]{Narayan98}, which means $\mdot < \sim
0.01\mdot_{\rm Edd}$, where $\mdot_{\rm Edd}$ is Eddington
accretion rate of RX J1856,
\bea \mdot_{\rm Edd}&=&4\pi m_{\rm p} cR/\sigma_{\rm T}\simeq
4.2\times10^{17} ~{\rm g~s^{-1}}, \eea
with $m_{\rm p}$ the proton mass and $\sigma_{\rm T}$ the Thomson
cross section.
The star should be in a propeller phase, and most of the accretion
matter can not be accreted onto the the star's surface (otherwise
the accretion induced X-ray luminosity will be much higher than
that we observed).

By the accretion flow, a quasi spherical layer may form at the
magnetospheric radius,
\be
r_{\rm m}\simeq({{B^2 R^6} \over { {\dot M}\sqrt{2GM} }})^{2/7}, %
\label{rm}
\ee
which is derived by equaling kinematic energy density of free-fall
matter to magnetic energy density. At the magnetospheric radius,
$\rmag$, accreted matter will decelerate and pile up to form a
boundary layer.
The density at the boundary layer should be high enough to make it
an optically thick region. The emission spectrum could then be
blackbody-like.

\subsection{ Constrain from temperature}

\begin{figure}
\plotone{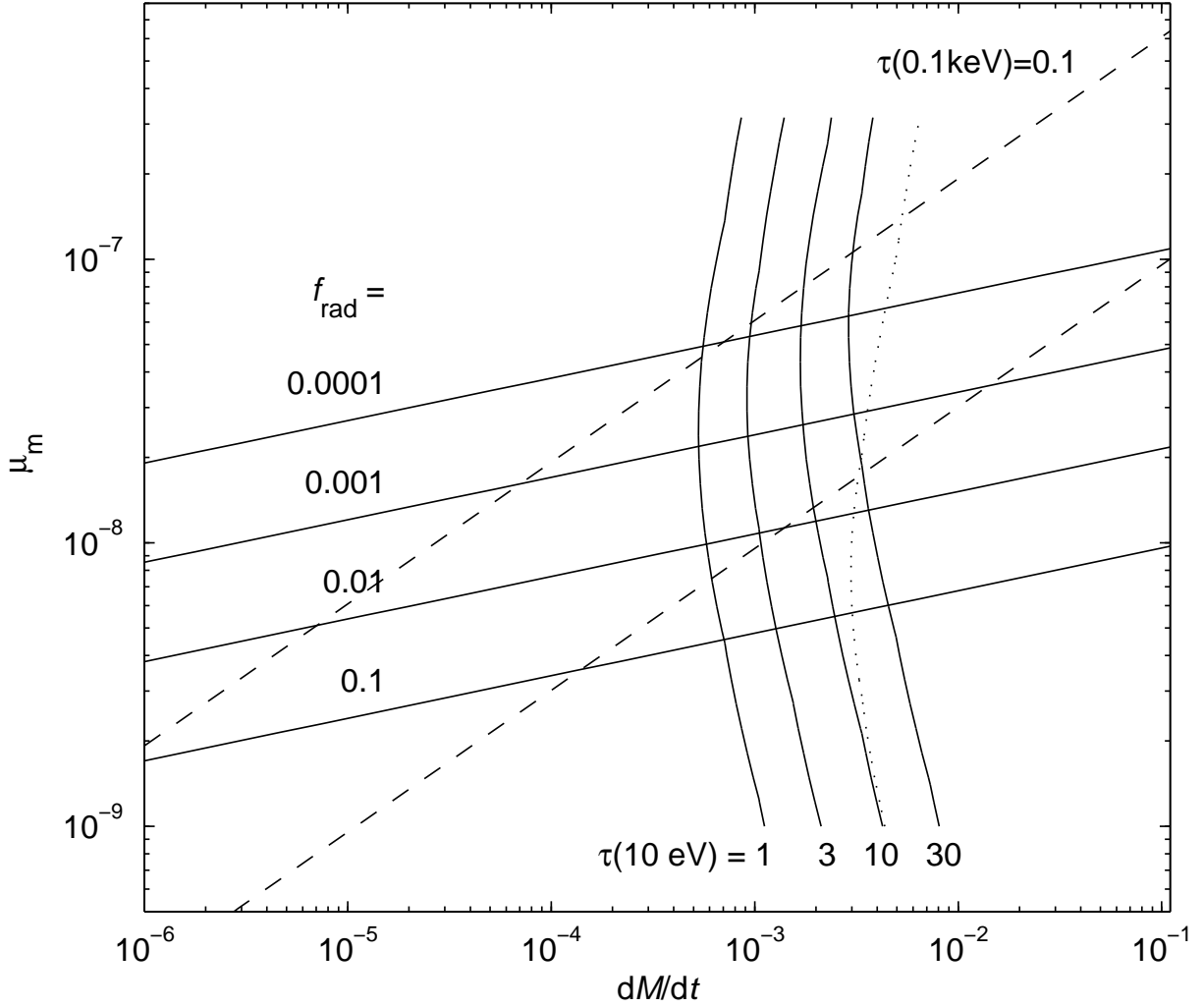} \figcaption{The magnetic moment per unit mass
$\mum$ (${\rm G ~cm^3 ~g^{-1}}$) vs. the accretion rate ${\rm
d}M/{\rm d}t$ (same to $\mdot$, in units of Eddington accretion
rate). The dashed line are constrains from Eq.(\ref{rm2}). Lines
labelled by different $\tau(10{\rm eV})$, the optical depth of the
layer for $h\nu = 10$ eV, are drawn via Eq.(\ref{kappaff}). The
dotted line is for $\tau(0.1{\rm keV)} = 0.1 $ with $h\nu = 0.1$
keV, through Eq.(\ref{kappaff}) too. Lines denoted by $\frad$ are
for different radiation efficiencies, from Eq.(\ref{frad}).
The allowed parameter space is limited by four lines: the two
dashed lines, the dotted line, and the line denoted by $\tau
(10{\rm eV})=1$.
It is evident that $8\times 10^{-9}\lesssim \mum({\rm G ~cm^3
~g^{-1}}) \lesssim 8\times10^{-8}$, $10^{-4} \lesssim \frad
\lesssim 10^{-2}$, and an accretion rate $10^{-3}\lesssim {\dot
M}/{\dot M}_{\rm Edd} \lesssim 10^{-2}$ which is consistent with
the ADAF assumption. \label{figct}}
\end{figure}

The UV-optical observation of RX J1856 is in the Rayleigh-Jeans
tail of the radiation from the boundary layer. The brightness of a
back body for $h\nu \ll kT$ is
$
B_\nu \simeq {2 k } \nu^2 T_{\rmag} /c^2
$
, where $k$ is Boltzman constant.
The observed flux per unit frequency should then be
\be F_\nu = \pi B_\nu \rmag^2/d^2 \simeq 2\pi  k { {\nu^2} \over
{c^2} }{ {1}\over {d^2}} T_{\rmag}\rmag^2, \ee
where $d$ is the the star's distance to the earth.
Observationally, one has
\be T_{\rmag}\rmag^2 = {\rm const. =  9.5\times 10^3 ~eV~km}^2.
\label{Trm} \ee
In order to be consistent with observation, the boundary layer's
temperature should be neither too high (not to affect the X-ray
spectrum) nor too low (to keep the Rayleigh-Jeans slope), i.e.
\citep{Burwitz03},
\be 4 ~{\rm eV} < T_{\rmag} <33 ~{\rm eV}. \label{<T>}\ee
From Eqs.(\ref{B-mum}), (\ref{rm}), (\ref{Trm}), and (\ref{<T>}),
one comes to
\be 17 ~{\rm km} < ({ {4M^2\mum^2} \over { {\dot M}\sqrt{2GM} }
})^{2/7} < 49 ~{\rm km}, \label{rm2} \ee
where only $\mum$ and $\mdot$ are free parameters if the star's
radius (and thus the mass through Eq.(\ref{M-R})) is determined
observationally. These two inequations of Eq.(\ref{rm2}) give two
dashed lines in $\mdot$-$\mum$ diagram and constrain the allowed
region to a small belt (see Fig. \ref{figct}).

\subsection{Constrain from optical depth}

To reproduce the observed spectrum in UV-optical bands, the
boundary layer should be optically thick.
We assume that the accreted matter's {\em radial} velocity
decreases from free fall velocity $v_{\rm ff}$ to zero at the
boundary layer. From mass continuity
$
\rho v = {{\mdot}/{4\pi \rmag^2} },
$
the density $\rho$ should increase inward.
The magnetic pressure at radius $r$ is $ \mathcal{P}(r) = {B(r)^2
/(8\pi)} = \mu^2/(2\pi r^6).$
Let's Consider a cubic mass with border $a$ and density $\rho$
near the boundary layer of the mass flow. The cube would feel a
force by the magnetic field
\be F \sim { {\partial \mathcal{P}} \over {\partial r}} a^3 . \ee
The pressure gradient ${\partial \mathcal{P}}/{\partial r}\propto
r_{\rm m}^{-7}$ is proximately a constant if the the layer is
thin. With the conservation of radial momentum, $-F{\rm d}t={\rm
d}(mv)$, and the definition of differential displacement, ${\rm
d}x=-v{\rm d}t$, we have $F{\rm d}x \sim mv{\rm d}v\sim a^3\rho v
{\rm d}v$. One could then estimate the $x$-value of the layer
($x\simeq 0$ at the bottom where $v\simeq 0$ and $\rho=\rho_{\rm
max}$) to be,
\be x \simeq v {{\mdot} \over {4\pi \rmag^2}} /{{\partial
\mathcal{P}} \over {\partial r} } \propto v . \ee
We obtain thus $\rho \propto {1/x}$ since $\rho v$ is a constant.
It would not be easy to estimate $\rho_{\rm max}$ since the
coupling between the accretion disk and the star's magnetosphere
is difficult to model. We have thus another free parameter,
$\rho_{\rm max}$, in our calculations.
Note that $v$ is the radial velocity. The total velocity does not
decrease so much because the velocity changes direction and most
of the accreted matter will be propelled out.

At temperature being higher than 4 eV, one can calculate from Saha
equation that nearly all hydrogen atoms is ionized. The opacity is
produced mainly by Thomson scattering, bond-free and free-free
absorptions. In the following, we show that free-free absorption
dominates.

\begin{figure}
\plotone{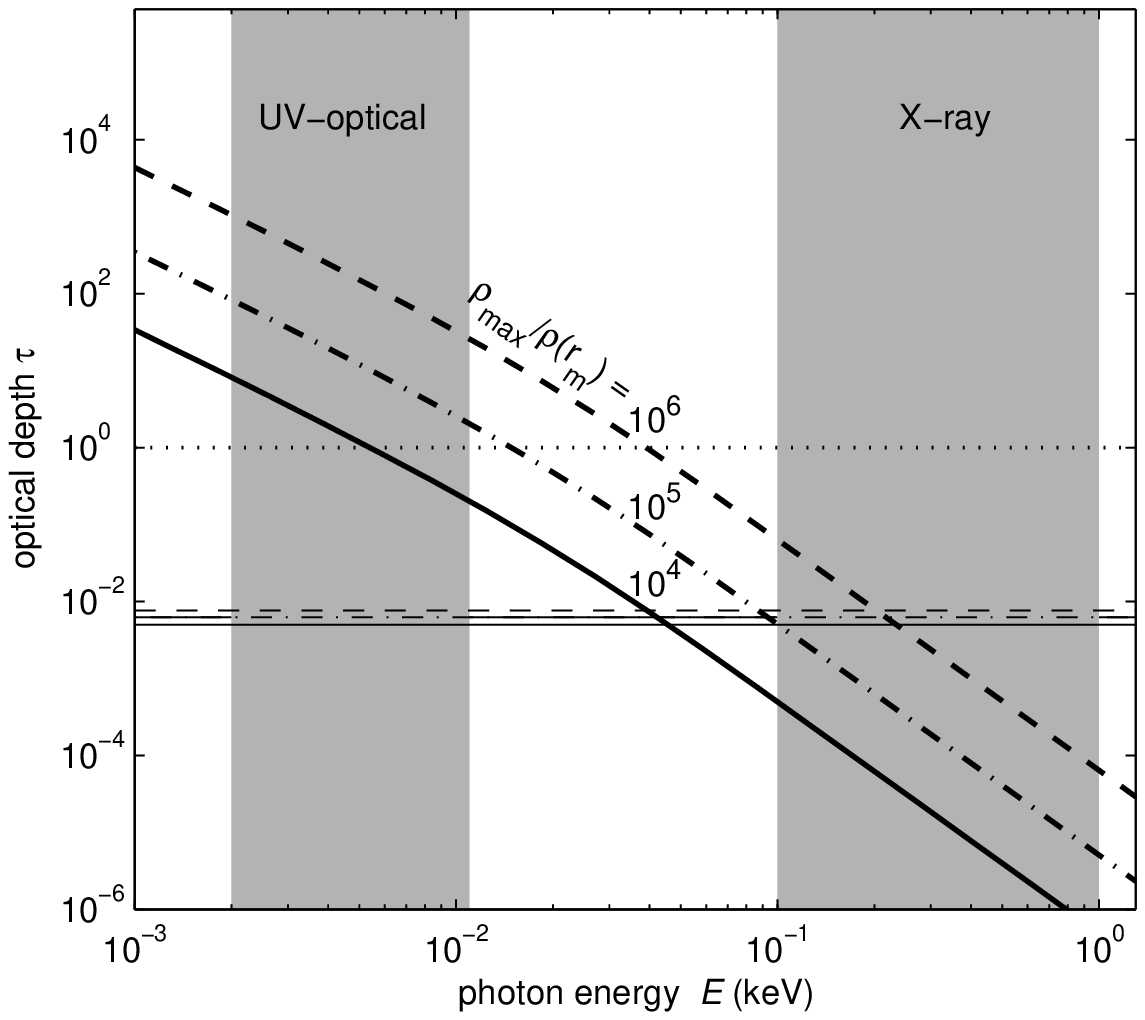} \figcaption{Plot of optical depth $\tau$ vs.
photon energy $E$. The bottom density, $\rho_{\rm max}$, is
considered as a free parameter.
Two hatched regions are for UV-optical and X-ray observations,
respectively. The dotted line is for $\tau=1$, beyond which the
layer could be optically thick. Thick lines are for free-free
absorption, while thin lines for Thomson scattering. Solid,
dash-dotted and dashed lines are for different $\rho_{\rm max}$.
It is found that the value $\rho_{\rm max} = (10^5\sim 10^6)\rho
(\rmag)$ in order to reproduce a boundary layer to be optically
thick in UV-optical but thin in X-ray bands, where $\rho
(\rmag)={\dot M}/(4\pi r_{\rm m}^2v_{\rm ff}(r_{\rm m}))$.
We choose ${\dot M}=10^{-3}{\dot M}_{\rm Edd}$ and $\mum=2\times
10^{-8}~{\rm G ~cm^3 ~g^{-1}}$ for indications here.
\label{figtau}}
\end{figure}

The Thomson cross section $\sigma_{\rm T}$ is a constant, and the
optical depth is very small if only Thomson scattering is
considered.
Free-free absorption is sensitive to density, so it is dominated
in the high density region and may result in an optically thick
boundary layer. Since free-free absorption coefficient \citep[p. 162]{Rybicki79}
\be \kappa_{\rm ff} = 3.7\times10^8T^{-1/2} Z^2 n_{\rm e} n_{\rm i} \nu^{-3}(1 - \exp(-\frac{h\nu}{kT})){\bar g}_{\rm ff}
, \label{kappaff} \ee
where $n_{\rm e}$ and $n_{\rm i}$ are electron and ion number
density, $Z\sim 1$ is ion charge and ${\bar g}_{\rm ff}\sim 1$ is
Gaunt factor, optical depth differs by a factor of $\sim 10^6$
between UV-optical and X-ray bands. Therefore it is possible that
the boundary layer would be always optically thin for the central
X-ray emission. Detail calculations confirm this suggestion (see
Fig. \ref{figtau}).

Bond-free and bond-bond absorptions should also be considered
because these processes may induce features in the spectrum which
are not observed yet. An estimation goes as following. For oxygen,
the most abundant element after hydrogen and helium, the typical
bond-free cross section $\sim10^{-19}$ cm$^2$ in 0.1--1 keV band
\citep{Reilman79}. Taking into account its abundance $\sim
10^{-3}$ (number relative to hydrogen), the effective cross
section is $\sigma_{\rm bb-bf}\sim 10^{-22}$ cm$^2$. The optical
depth induced by these absorptions could be smaller than 1 though
$\sigma_{\rm bb-bf}$ is about $\sim 10^2$ times that of electrons.
The model proposed may then still work.

\section{The results}

\subsection{The radiation efficiency in the layer}

Since the star is in a propeller phase, most of the accreted
matter will be expelled out finally. In this way only a small part
of the accretion energy may be converted to radiation. Here we use
$\frad$ to represent the transformation rate. We have then
\be { {G M \mdot} \over {\rmag} } \frad = \sigma T^4 \obl \rmag^2
, \label{gra2T} \ee
where $\obl$ is the emission angle of the boundary layer which is
$4\pi$ if the boundary layer is completely spherical. Since
$\frad$ and $\obl$ are degenerate, we adopt simply $\obl=4\pi$ in
the following.

Combining with Eqs.(\ref{B-mum}), (\ref{rm}), (\ref{Trm}), and
(\ref{gra2T}), one may obtain the value of a new free parameter
$\frad$, the dependence of which on other two free parameters is
\be \frad \simeq 9.24\times10^{-32} { {\mdot^{3/7}} \over {\mum^{20/7}} }.
\label{frad} \ee
Different $\frad$ value gives different lines on $\mdot$-$\mum$
diagram (see Fig. \ref{figct}). The allowed range of $\frad$
should be about $10^{-2}\sim 10^{-4}$.

\subsection{The star's spin period and polar magnetic field}

\begin{figure}
\plotone{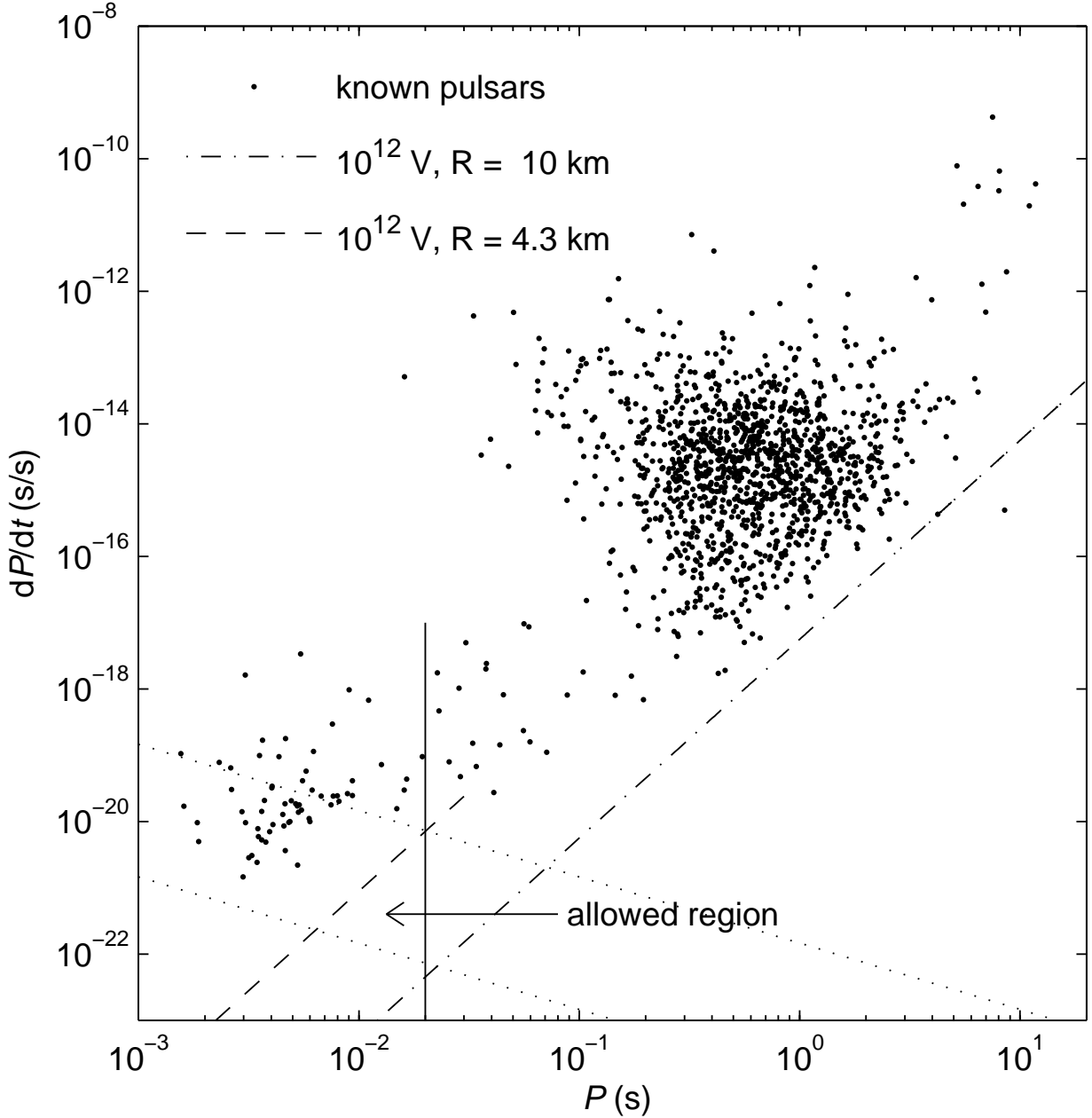} \figcaption{%
The $P-{\dot P}$ diagram and the possible position of RX J1856. RX
J1856 should be in the left of the solid vertical line because of
$P<20$ ms.
It should also be in-between two dotted lines through $4\times10^7
\lesssim B {\rm G} \lesssim 4\times10^8 {\rm G}$. The dash-dotted
and dashed lines are approximately death lines for normal pulsar
with radius $R=10$ km and for RX J1856 with $R=4.3$ km,
respectively.
It is suggested that RX J1856 could be in the triangle (i.e., the
``{\em allowed region}'') if the star spins down due to pure
magnetodipole radiation.
No magnetospheric activity exists on the star because of a low
potential drop in the open field line region. The pulsar data are
from ATNF Pulsar Catalogue
\citep[http://www.atnf.csiro.au/research/pulsar/psrcat]{Manchester05}.
\label{fig3}}
\end{figure}

A propeller phase requires $\rco  < \rmag$, where
$\rco=(GMP^2/(4\pi^2))^{1/3} \simeq 6.88\times10^7 P^{2/3} $ cm is
the corotating radius. We can thus constrain the star's spin
period to be $ \lesssim 20$ ms from this requirement. On the
$P-{\dot P}$ diagram, a death line may simply be represented by an
equal voltage line \citep{zhang03}.
We therefore constrain $P$ and $\dot P$ of RX J1856 via setting
the death voltage to be $10^{12}$ V in order to produce
approximately the dash-dotted death line for normal pulsars in
Fig.3. The death line would be described by
\be \Phi \simeq 6.6\times 10^{12} B_{12} P^{-2} R_6^3  ~~{\rm V}
\simeq 10^{12}~~{\rm V}, \ee
where $B_{12}$ is the magnetic field strength in units of
$10^{12}$ G and $R_6$ is star radius in units of $10^6$ cm. The
line for RX J1856 is then $0.5 B_{12}P^{-2}\simeq 1$.
Note that since low-mass quark stars could have small radii, their
death line would be different from that of the normal pulsars.
A constrain for magnetic polar field \citep{xu05}, $B=32\pi{\bar
B}\mum/3$, is obtained so as to make the star radio quiet,
$4\times10^7 {\rm G} \lesssim B \lesssim 4\times10^8 {\rm G}$,
from the limit $8\times 10^{-9}\lesssim \mum({\rm G ~cm^3
~g^{-1}}) \lesssim 8\times10^{-8}$.
We could thus obtain $10^{-22}\lesssim \pdot \lesssim 10^{-20}$ in
Fig. \ref{fig3} if a spindown torque by pure magnetodipole
radiation is assumed.

\section{Conclusions and discussions}

We propose that RX J1856.5-3754 is a low-mass quark star, which is
in a propeller phase.
A boundary layer of RX J1856 is suggested and modelled in order to
understand the UV-optical excess observed. In order that the layer
could be optically thin in X-ray but thick in UV-optical bands,
some constrains for RX J1856 are obtained: the polar magnetic
field $4\times10^7 {\rm G} \lesssim B \lesssim 4\times10^8 {\rm
G}$, the spin period $P\la 20$ ms, and an accretion rate
$10^{-3}\lesssim {\dot M}/{\dot M}_{\rm Edd} \lesssim 10^{-2}$, if
the the star's radius and mass are 4.3 km and $\sim 0.1M_\odot$,
respectively.
Additionally, we find that the radiation efficiency of accretion
matter in the layer could be between $10^{-2}$ and $10^{-4}$, and
that the density at the layer's bottom is about $10^{5\sim 6}$
times that at the top.
No magnetospheric activity exists on the star since it would be
under its death line, which could be the reason that neither radio
emission nor X-ray modulation is observed from RX J1856.
The star could have a weak magnetic field to be similar to that of
the millisecond pulsars.

In order to reproduce the UV-optical spectrum, an optically thick
layer is considered only, which results naturally a slope of the
Rayleigh-Jeans tail.
We are still not sure whether the UV-optical spectrum observed
could also be understood by an optically thin boundary layer. The
difficulty in the later case could be to calculate the emissivity
of matter with strong magnetic field.
Even in the first case of an optically thick layer, the radiation
efficiency is still hard to calculate due to few knowledge of the
interaction between charged particles and magnetic field.
We obtain a low radiation efficiency, $10^{-4\sim -2}$, in our
constrains, which could be reasonable since most of the kinematic
energy is take out during a propeller process. The emission from
the outer part of the ADAF disk is not taken into account since
the luminosity should be very low, with emission in low-energy
band.

The bag constant (or effectively, the average density of the star
$4{\bar B}$) is chosen to be $\bbag = 1.5\times10^{14}~ {\rm
g~cm^{-3}}$ in our calculations.
However, the allowed region in $P-{\dot P}$ diagram will be larger
if $\bar B$ is released from 1.07$\times10^{14}~ {\rm g~cm^{-3}}$
to 1.96$\times10^{14}~ {\rm g~cm^{-3}}$.
The model proposed in this paper would thus be a flexible one.

How to check the model proposed?
A direct way might be to find the star's spin period in radio
and/or X-ray data. This model would be rule out if the spin period
is much larger than 20ms.
Photons in low-energy bands (e.g., infrared or submillimeter
emission) would also be radiated from the ADAF disk and the
boundary layer. The luminosity of this emission could be high
enough if more powerful facilities are offered in the future.

\acknowledgments

We would like to acknowledge Dr. Fukun Liu for his helpful
discussions. This work is supported by NSFC (10273001), the
Special Funds for Major State Basic Research Projects of China
(G2000077602), and by the Key Grant Project of Chinese Ministry of
Education (305001).

\clearpage

\end{document}